\documentclass[12pt]{iopart}

\usepackage{graphicx}

\begin{document}

\title{Is there Quark Matter in (Low-Mass) Pulsars?}

\author{Irina Sagert, Giuseppe Pagliara, Matthias Hempel, and
J\"urgen Schaffner--Bielich\footnote{talk given at the QM2008
 conference, Jaipur, India, February 4--10, 2008}}

\address{Institut f\"ur Theoretische Physik/Astrophysik,
J. W. Goethe Universit\"at,
D--60054~Frankfurt am Main, Germany}

\ead{schaffner@astro.uni-frankfurt.de}

\begin{abstract} 
  The effect of the QCD phase transition is studied for the mass-radius
  relation of compact stars and for hot and dense matter at a given
  proton fraction used as input in core-collapse supernova simulations.
  The phase transitions to the 2SC and CFL color superconducting phases
  lead to stable hybrid star configurations with a pure quark matter
  core. In supernova explosions quark matter could be easily produced
  due to $\beta$-equilibrium, small proton fractions and nonvanishing
  temperatures. A low critical density for the phase transition to quark
  matter is compatible with present pulsar mass measurements.
\end{abstract}

The nuclear equation of state serves as a crucial input for several
astrophysical systems. In supernovae simulations temperatures of
$T=20$~MeV and higher are reached with a baryon number density slightly
above normal nuclear matter density at bounce \cite{Liebendoerfer:2007hq}. 
In the subsequent evolution of the hot proto-neutron star the central
density increases to several times normal nuclear matter density. In
neutron star mergers the temperatures and densities reached are also in
ranges where one expects the QCD phase transition from hadronic matter
to a chirally restored phase.
Hunting down strange quark matter in the heavens is coming of age and
several signals have been suggested (see \cite{SchaffnerBielich:2007mr}
for a review) as exotic mass-radius relations of compact stars, rapidly
rotating pulsars due to a r-mode {\em stability} window, enhanced cooling
of neutron stars, gamma-ray bursts by transition to strange quark
matter, and gravitational wave signals from phase transitions in binary
neutron star collisions, in the collapse of neutron star to a hybrid star or
in the r-mode spin-down of hybrid stars. Interestingly, core-collapse
supernovae have not been studied in that detail with regard to the QCD
phase transition.
One recent analysis in connection with quark stars has been done on
the x-ray burster EXO 0748-676 stating that the mass and the
radius should be in the range $M\geq 2.10\pm 0.28 M_\odot$ and $R\geq
13.8 \pm 1.8$ km \cite{Ozel:2006km}. Contrary to the claims made by the
author of \cite{Ozel:2006km}, these constraints do not rule out quark
stars or hybrid stars but rather soft equations of state
\cite{Alford:2006vz}. 
In addition, a multiwavelength analysis arrives at the conclusion that the
data is more consistent with a mass of $M=1.35M_\odot$
\cite{Pearson:2006zy}. Also, the mass of the pulsar J0751+1807 has been
corrected 
from $(2.1\pm 0.2)M_\odot$ down to a value of $(1.26\pm
0.14)M_\odot$ recently \cite{Nice:2008}. We point out that the masses of
the supermassive pulsars reported in \cite{Freire:2007jd} are not real
mass measurements as they rely on a statistical analysis of the
orbital inclination angle of the binary pulsar system. We
conclude that the present firm mass limit for neutron stars is still
given by the Hulse-Taylor pulsar mass of $1.44M_\odot$.

\begin{figure}
\centerline{\includegraphics[width=0.54\textwidth]{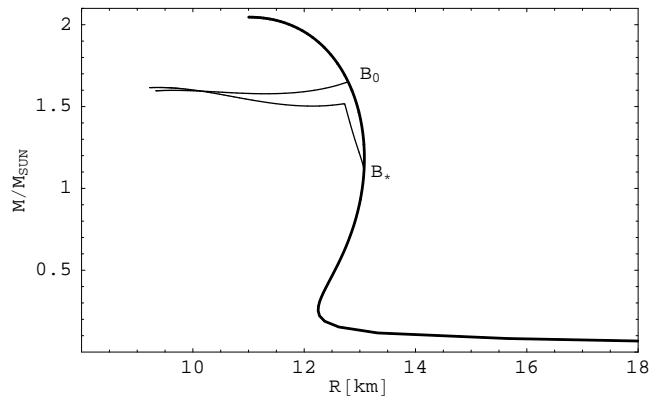}}
\caption{The mass-radius plot for hybrid stars showing the stable
  branches with a color-superconducting quark core (taken from
  \cite{Pagliara:2007ph}).}
\label{fig:mr_csc}
\end{figure}

There might be even more than one phase transition at high densities as
there are several phases of color superconducting quark matter in
$\beta$ equilibrium, like the two-flavor color superconducting phase (2SC),
the color-flavor locked phase (CFL) and its gapless versions, see
\cite{Ruster:2005jc,Sandin:2007zr}.
The generic impact on the global properties of compact stars is that a
phase transition close to the maximum mass configuration will make the
solutions with a quark core unstable \cite{Pagliara:2007ph}. A
smaller critical energy density for the phase transition can generate a
stable quark core. Within the Nambu--Jona-Lasinio model both the
dynamical quark masses (quark 
condensates) and the color-superconducting gaps $\Delta$ are described.
The 't Hooft term and a vector interaction term are taken into account.
There are two free parameters left, one describing the diquark coupling
$G_D$ and another one the strength of the repulsive vector
interaction $G_V$. In addition, the vacuum pressure is fixed either in
vacuum ($B_0$) or at the chiral phase transition ($B_*$). In the latter
case the switch from the hadronic phase (using a
relativistic mean field approach) to the NJL quark description is
exactly happening at the chiral phase transition. The difference in the
two vacuum pressures is small but significantly moves the phase
transition point to lower baryon densities.
The hybrid star mass-radius diagram is depicted in
Fig.~\ref{fig:mr_csc}. In the first case ($B_0$), the phase transition
produces directly the CFL phase at high densities. The corresponding
compact star is unstable first but is then stabilised for larger central
densities. In the second case ($B_*$), two phase transitions are present,
one from hadronic matter to the 2SC phase and the other one from the 2SC phase
to the CFL phase. The two phase transitions produce two kinks in the
mass-radius curve with two branches of stable solutions. The new stable
solution constitutes the third family of compact stars.
These phase transitions might have also implications for
core-collapse supernova explosions by producing a secondary shock wave
or by modifying the gravitational wave pattern from coalescing neutron
stars.

\begin{figure}
\centerline{
\includegraphics[angle=-90,width=0.5\textwidth]{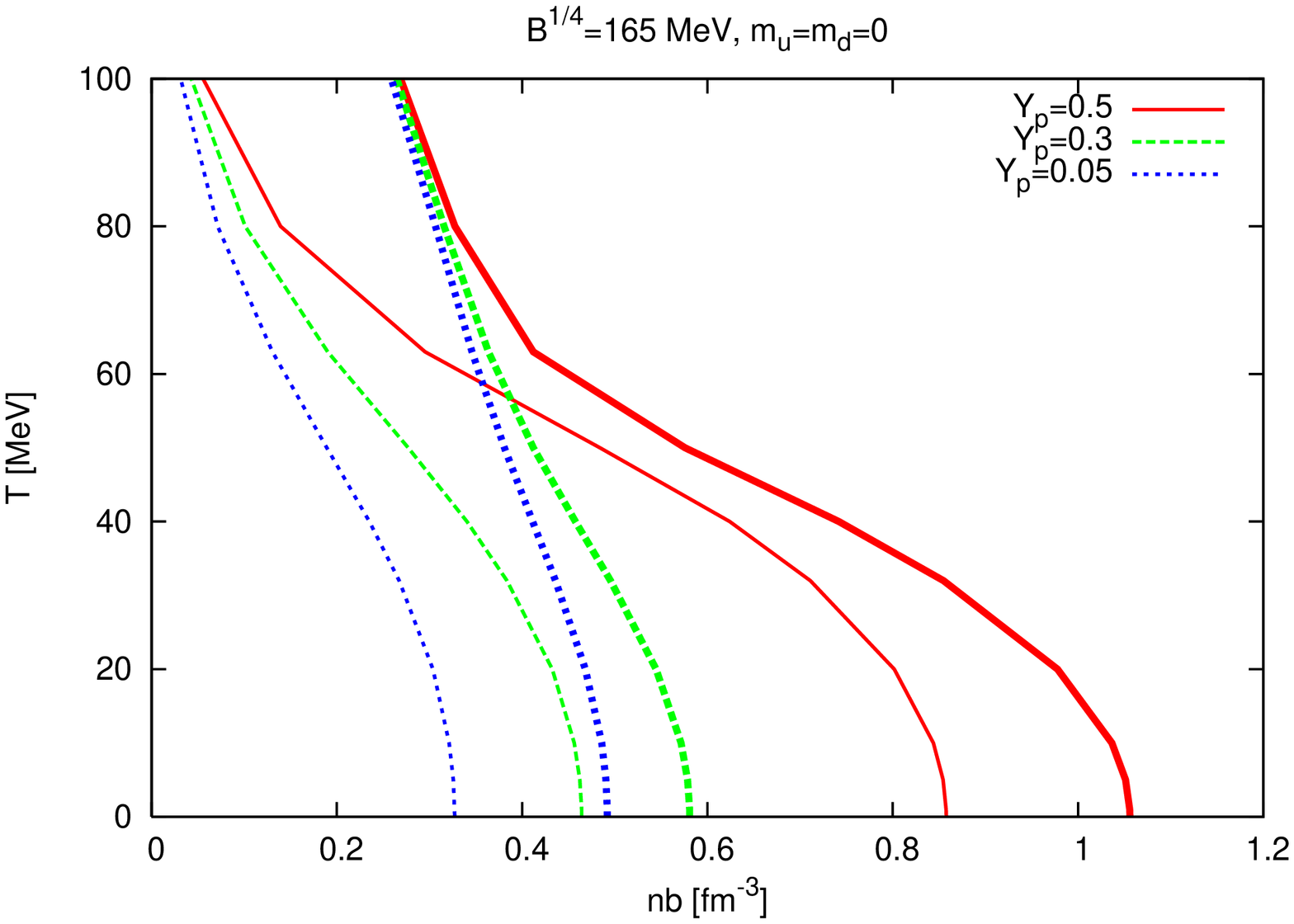}
\includegraphics[angle=-90,width=0.5\textwidth]{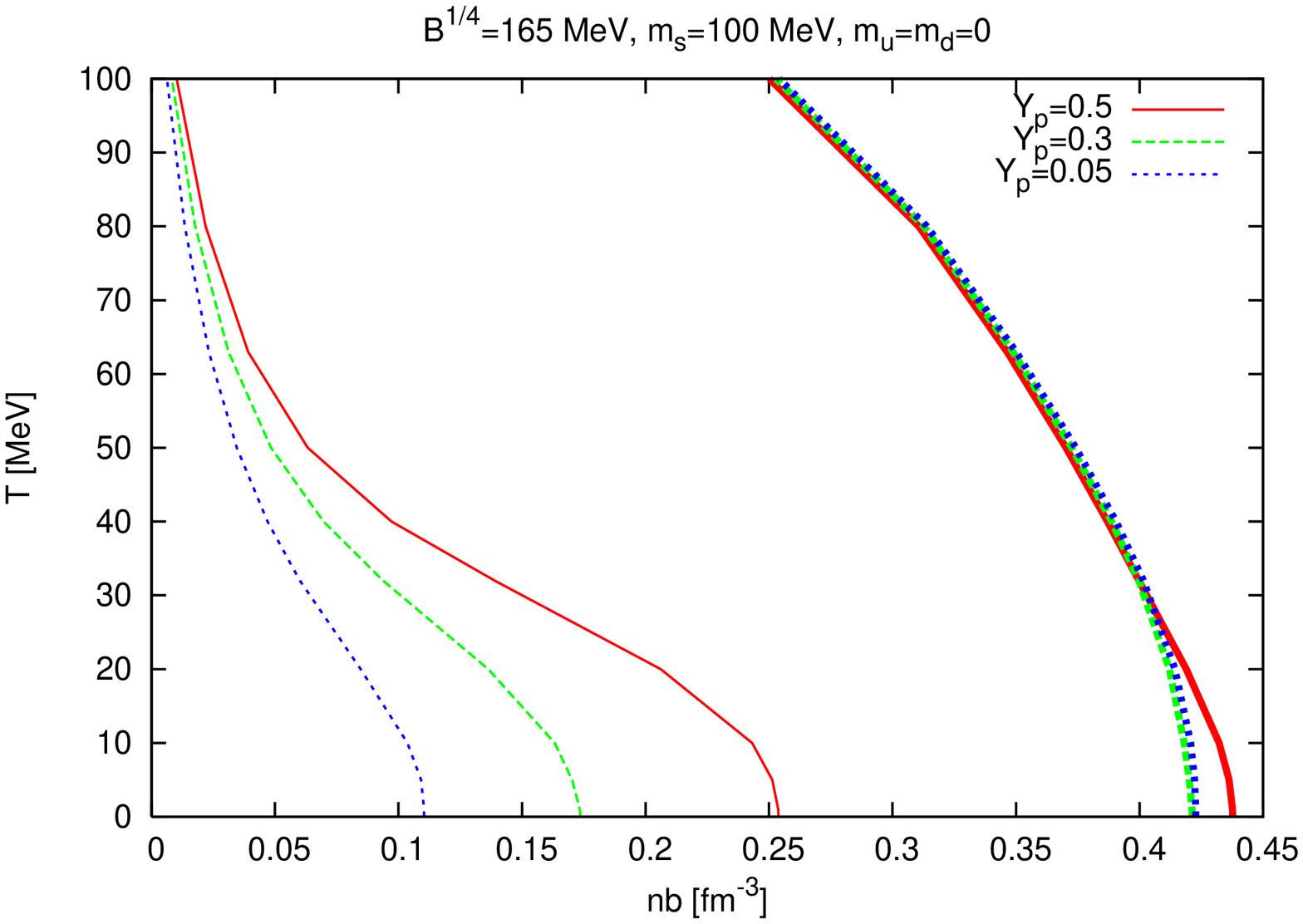}}
\caption{The phase transition lines to quark matter using the MIT bag
  constant for conditions relevant for heavy-ion collisions (left plot)
  and supernova explosions (right plot). Shown is the onset (thin lines)
  and the end of the mixed phase (thick lines).}
\label{fig:phasetrans_hic}
\end{figure}

In heavy-ion collisions the phase transition density to quark matter is
different from supernova explosions. In the former there is no
$\beta$-equilibrium established and consequently the pure hadronic and
pure quark phases have vanishing net strangeness initially.
Figure~\ref{fig:phasetrans_hic} shows the phase
transition line between hadronic (the relativistic
mean field model TM1) and a MIT bag equation of state at nonvanishing
temperature and baryon density. Just up-
and down-quarks are considered in the quark phase as there is no
$\beta$-equilibrium. The critical density for the phase transition is
quite large in particular for isospin-symmetric matter (proton fraction
of $Y_p=0.5$). Hence, the production of ud-quark matter is unfavoured
for heavy-ion collisions at small temperatures and high proton
fractions. The phase transition at large chemical
potentials for ud-quark matter is located at higher values than the extracted
freeze-out parameters from the statistical analysis of
particle ratios at SIS and AGS energies.
The phase transition to quark matter in astrophysical
environments is dictated by $\beta$-equilibrium. The start of the
mixed phase occurs already at quite low densities based on the enhanced
stability of strange quark matter compared to ud-quark matter. The
critical densities for isospin-asymmetric matter (low proton fraction
$Y_p$) are even smaller due to the large asymmetry energy of nuclear
matter. In comparison to cold neutron star matter, hot supernova
environments favour the presence of quark matter as the phase transition
line bends 
towards low chemical potentials for large temperatures. These three
effects combined favour the production of quark matter already at quite
early stages of the supernova evolution. We note that at supernova
bounce typically $T\approx 20$~MeV and $Y_p\approx 0.3$ with densities
slightly above saturation density $n_0=0.15$~fm$^{-3}$, so that quark
matter could be present already at supernova bounce!
The early onset of the quark matter phase corresponding to low values of
the MIT bag constant was rejected in previous work with the argument
that it produces a too small maximum mass for cold neutron stars
\cite{Pons:2001ar}. This statement is not entirely correct, as 
small values of the bag constant result actually in a larger maximum
mass {\em if} quark matter is the dominant phase of the compact star. We find
that with the quark equation of state used here the maximum mass is well
above the 1.44~$M_\odot$ limit. The presence of quark matter drastically
changes the mass-radius diagram. A stable solution
appears with rather small radii for low-mass 
compact stars serving as a unique feature for the presence of exotic matter.

In summary, the phase transition to quark matter can have an enormous
impact on the physics of compact stars and supernova explosions as the
quark matter phase transition line is located at lower densities
compared to the environment created in heavy-ion experiments. Hunting
for supermassive neutron 
stars is, however, not a way to pin down the presence of quark matter in
the heavens. Masses of neutron stars well above $2.3 M_\odot$ are hard
to achieve with {\em any} realistic modern nuclear equation of state.
Constraints from heavy-ion data points towards a soft nuclear equation
of state at moderate densities so that extremely large maximum masses
are unfavoured \cite{Sagert:2007nt}. A maximum mass of around $2M_\odot$
can be a compact star consisting just of normal nuclear matter or a
hybrid star with a quark matter core \cite{Alford:2006vz}. If the maximum
mass turns out to be about $1.5M_\odot$ (well below $2M_\odot$) exotic
matter is involved which lowers the maximum mass, be it hyperons, kaon
condensates or quarks. By measuring the radius of light pulsars, e.g.\
with masses of 1.2~$M_\odot$, the equation of state and the possible
presence of exotic matter could be revealed with less ambiguities. If
quark matter is present already at quite low densities it will lower
drastically the radius of light compact stars. The conditions in core
collapse supernovae are particularly favourable for the onset of
the quark matter phase.

I.S. gratefully acknowledges support from the Helmholtz Research School
for Quark Matter Studies, G.P. from INFN and M.H. from the Frankfurt
Institute for Advanced Studies. J.S.B. thanks the Deutsche
Forschungsgemeinschaft for their support to participate at the QM2008
conference.

\section*{References}

\bibliographystyle{revtex}
\bibliography{all,literat}

\begin{thebibliography}{10}
\providecommand*{\bibinfo}[2]{#2}
\providecommand*{\eprint}[1]{#1}
\providecommand*{\url}[1]{#1}
\bibitem{Liebendoerfer:2007hq}
\bibinfo{author}{M.~Liebend{\"o}rfer}, \bibinfo{author}{T.~Fischer},
  \bibinfo{author}{C.~Fr{\"o}hlich}, \bibinfo{author}{F.-K. Thielemann}, and
  \bibinfo{author}{S.~Whitehouse}, \bibinfo{journal}{J. Phys. G}
  \bibinfo{volume}{\textbf{35}}, \bibinfo{pages}{014056}
  (\bibinfo{date}{2008}), \eprint{arXiv:0708.4296 [astro-ph]}.
\bibitem{SchaffnerBielich:2007mr}
\bibinfo{author}{J.~Schaffner-Bielich}, \bibinfo{journal}{PoS}
  \bibinfo{volume}{\textbf{(CPOD2007)}}, \bibinfo{pages}{062}
  (\bibinfo{date}{2007}), \eprint{arXiv:0709.1043 [astro-ph]}.
\bibitem{Ozel:2006km}
\bibinfo{author}{F.~{\"O}zel}, \bibinfo{journal}{Nature}
  \bibinfo{volume}{\textbf{441}}, \bibinfo{pages}{1115} (\bibinfo{date}{2006}),
  \eprint{astro-ph/0605106}.
\bibitem{Alford:2006vz}
\bibinfo{author}{M.~Alford}, \bibinfo{author}{D.~Blaschke},
  \bibinfo{author}{A.~Drago}, \bibinfo{author}{T.~Kl\"ahn},
  \bibinfo{author}{G.~Pagliara}, and \bibinfo{author}{J.~Schaffner-Bielich},
  \bibinfo{journal}{Nature} \bibinfo{volume}{\textbf{445}}, \bibinfo{pages}{E7}
  (\bibinfo{date}{2006}), \eprint{astro-ph/0606524}.
\bibitem{Pearson:2006zy}
\bibinfo{author}{K.~J. Pearson}, \bibinfo{author}{R.~Hynes},
  \bibinfo{author}{D.~Steeghs}, \bibinfo{author}{P.~Jonker},
  \bibinfo{author}{C.~Haswell}, \bibinfo{author}{A.~King},
  \bibinfo{author}{K.~O'Brien}, \bibinfo{author}{G.~Nelemans}, and
  \bibinfo{author}{M.~Mendez}, \bibinfo{journal}{Astrophys. J.}
  \bibinfo{volume}{\textbf{648}}, \bibinfo{pages}{1169} (\bibinfo{date}{2006}),
  \eprint{astro-ph/0605634}.
\bibitem{Nice:2008}
\bibinfo{author}{D.~J. {Nice}}, \bibinfo{author}{I.~H. {Stairs}}, and
  \bibinfo{author}{L.~E. {Kasian}}, \bibinfo{journal}{AIP Conference
  Proceedings} \bibinfo{volume}{\textbf{983}}, \bibinfo{pages}{453}
  (\bibinfo{date}{2008}).
\bibitem{Freire:2007jd}
\bibinfo{author}{P.~C.~C. Freire}, \bibinfo{author}{S.~M. Ransom},
  \bibinfo{author}{S.~Begin}, \bibinfo{author}{I.~H. Stairs},
  \bibinfo{author}{J.~W.~T. Hessels}, \bibinfo{author}{L.~H. Frey}, and
  \bibinfo{author}{F.~Camilo}, \bibinfo{journal}{arXiv:0711.0925 [astro-ph]}
  (\bibinfo{date}{2007}).
\bibitem{Pagliara:2007ph}
\bibinfo{author}{G.~Pagliara} and \bibinfo{author}{J.~Schaffner-Bielich},
  \bibinfo{journal}{Phys. Rev. D} \bibinfo{volume}{\textbf{77}},
  \bibinfo{pages}{063004} (\bibinfo{date}{2008}), \eprint{arXiv:0711.1119
  [astro-ph]}.
\bibitem{Ruster:2005jc}
\bibinfo{author}{S.~B. R{\"u}ster}, \bibinfo{author}{V.~Werth},
  \bibinfo{author}{M.~Buballa}, \bibinfo{author}{I.~A. Shovkovy}, and
  \bibinfo{author}{D.~H. Rischke}, \bibinfo{journal}{Phys. Rev. D}
  \bibinfo{volume}{\textbf{72}}, \bibinfo{pages}{034004}
  (\bibinfo{date}{2005}), \eprint{hep-ph/0503184}.
\bibitem{Sandin:2007zr}
\bibinfo{author}{F.~Sandin} and \bibinfo{author}{D.~Blaschke},
  \bibinfo{journal}{Phys. Rev. D} \bibinfo{volume}{\textbf{75}},
  \bibinfo{pages}{125013} (\bibinfo{date}{2007}), \eprint{astro-ph/0701772}.
\bibitem{Pons:2001ar}
\bibinfo{author}{J.~A. Pons}, \bibinfo{author}{A.~W. Steiner},
  \bibinfo{author}{M.~Prakash}, and \bibinfo{author}{J.~M. Lattimer},
  \bibinfo{journal}{Phys. Rev. Lett.} \bibinfo{volume}{\textbf{86}},
  \bibinfo{pages}{5223} (\bibinfo{date}{2001}), \eprint{astro-ph/0102015}.
\bibitem{Sagert:2007nt}
\bibinfo{author}{I.~Sagert}, \bibinfo{author}{M.~Wietoska},
  \bibinfo{author}{J.~Schaffner-Bielich}, and \bibinfo{author}{C.~Sturm},
  \bibinfo{journal}{J. Phys. G} \bibinfo{volume}{\textbf{35}},
  \bibinfo{pages}{014053} (\bibinfo{date}{2008}), \eprint{arXiv:0708.2810
  [astro-ph]}.

\end{thebibliography}

\end{document}